# Precise Fabrication of Uniform Molecular Gaps for Active Nanoscale Devices

Farnaz Niroui,[1]* Mayuran Saravanapavanantham,[1]† Jinchi Han,[1]† Jatin J. Patil,[2] Timothy M. Swager,[3] Jeffrey H. Lang,[1]* Vladimir Bulović[1]*

**Affiliations**

[1]Department of Electrical Engineering and Computer Science, Massachusetts Institute of Technology, Cambridge, MA 02139, USA

[2]Department of Material Science and Engineering, Massachusetts Institute of Technology, Cambridge, MA 02139, USA.

[3]Department of Chemistry, Massachusetts Institute of Technology, Cambridge, MA 02139, USA.

†These authors contributed equally to this work.
*Corresponding authors. Email: fniroui@mit.edu, lang@mit.edu, bulovic@mit.edu

**Abstract**

Molecules with versatile functionalities and well-defined structures, can serve as building blocks for extreme nanoscale devices. This requires their precise integration into functional heterojunctions, most commonly in the form of metal-molecule-metal architectures. Structural damage and nonuniformities caused by current fabrication techniques, however, limit their effective incorporation. Here, we present a hybrid fabrication approach enabling uniform molecular gaps. Template-stripped lithographically-patterned gold electrodes with sub-nanometer roughness are used as the bottom contacts upon which the molecular layer is formed through self-assembly. The top contacts are assembled using dielectrophoretic trapping of colloidal gold nanorods, resulting in uniform sub-5 nm junctions. In these electrically-active designs, we further explore the possibility of mechanical tunability. The presence of molecules may help control sub-nanometer mechanical modulation which is conventionally difficult to achieve due to instabilities caused by surface adhesive forces. Our approach is versatile, providing a platform to develop and study active molecular gaps towards functional nanodevices.

## Introduction

As device miniaturization reaches the few-nanometer dimensions, molecules emerge as attractive building blocks to enable novel nanoscale devices. Once studied primarily for their electronic transport properties, the versatile functionalities of molecules have extended their prospects to diverse applications in electronics, optics, mechanics, thermoelectrics and spintronics (1-4). Moreover, molecules whose structures can be precisely designed through chemical synthesis and assembled into well-defined ensembles have been effective components to help define heterostructures with feature sizes beyond the resolution limit of conventional fabrication techniques. The resulting small dimensions give access to unique phenomena inherent to their unprecedented size, which can be further complemented by molecular functionalities (5-7). Collectively, a platform is introduced that can empower next-generation functional nanodevices with capabilities beyond conventional technologies.

However, given their ultimate small dimensions, integration of molecules into functional structures which commonly take the form of metal-molecule-metal motifs is not trivial. Common top-down fabrication strategies lack the necessary resolution, introduce nonuniformities on the order of the molecular dimensions, and induce structural damage. These shortcomings detrimentally influence device functionality. Such effects are exacerbated when the molecular layer is not tightly packed or thermally stable. On the other hand, bottom-up assembly strategies combined with self-assembly capabilities of the molecules can form well-defined molecular gaps. However, these structures generally lack the complexity to accommodate effective incorporation of contacts needed for electrical probing in an active operation. To this end, various fabrication techniques have continuously been developed (3, 4, 8-15). An optimal approach should yield a well-defined molecular layer sandwiched between two contacts, each with sub-nanometer surface uniformity, while allowing electrical access. The technique should also accommodate diverse material systems and large area processability as needed.

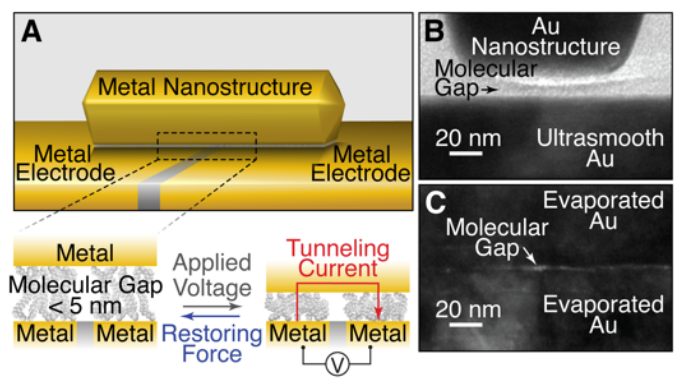

*Fig. 1. A uniform electrically-active molecular gap. (A) Schematic of a device composed of a nanorod bridging ultrasmooth bottom contacts separated by a self-assembled molecular layer. The mechanically mobile design of the top contact may enable gap reconfiguration during actuation. (B) Transmission electron microscope (TEM) cross-section image of a uniform molecular junction formed between an ultrasmooth gold (Au) thin-film and an atomically-flat facet of a gold nanorod. (C) TEM cross-section image of a nonuniform molecular junction formed between thermally evaporated gold contacts.*

Here, we present a hybrid technique for room temperature fabrication of uniform, electrically-active, < 5 nm molecular gaps. Lithographically patterned electrodes and wires are template-stripped to form the bottom contacts and desired circuitry in a scalable manner, while maintaining sub-nanometer surface uniformity. Molecules are subsequently self-assembled with the top contact introduced through dielectrophoretic trapping of colloidal gold nanorods, bridging the two laterally-spaced bottom electrodes, as shown in Fig. 1a. The result is a uniform metal-molecule-metal junction with gap separation precisely defined based on the thickness of the self-assembled layer. This is in contrast to what is feasible through commonly employed metal deposition techniques such as thermal evaporation, where surface roughness and metal penetration prevent a well-defined structure (Fig. 1b and c). Unique to this design, the top contact can be mechanically active, responding to externally-applied stimuli, such as an electrostatically induced force using the bottom contacts. Given the highly sensitive structure-property relations at these small nanometer dimensions, structural reconfiguration of the molecular gap can introduce new means of active device engineering. In this work, we demonstrate the versatility of our fabrication technique and explore the possibility of mechanical tunability by studying nanogaps with self-assembled molecules expected to exhibit different morphologies and mechanical properties.

**Results**

**Ultrasmooth metal surfaces for large area fabrication of bottom contacts**

An ultrasmooth bottom contact is critical in facilitating uniform self-assembly of the molecules and the formation of a well-defined molecular gap (16, 17). Commonly used metals, as deposited, have roughness and nonuniformities on the order of the molecular dimensions, limiting their use for direct integration into such junctions. Likewise, lithographic techniques used in patterning of these contacts often result in edge imperfections, artefacts inherent to the processing. Unlike most applications where such small defects are considered insignificant, for applications where critical dimensions are on the order of the molecule sizes, even minute nonuniformities can translate into drastic structural and consequently functional variability and instability. To this end, we implement a template-stripping technique (16-18) following the steps outlined in Fig. 2a.

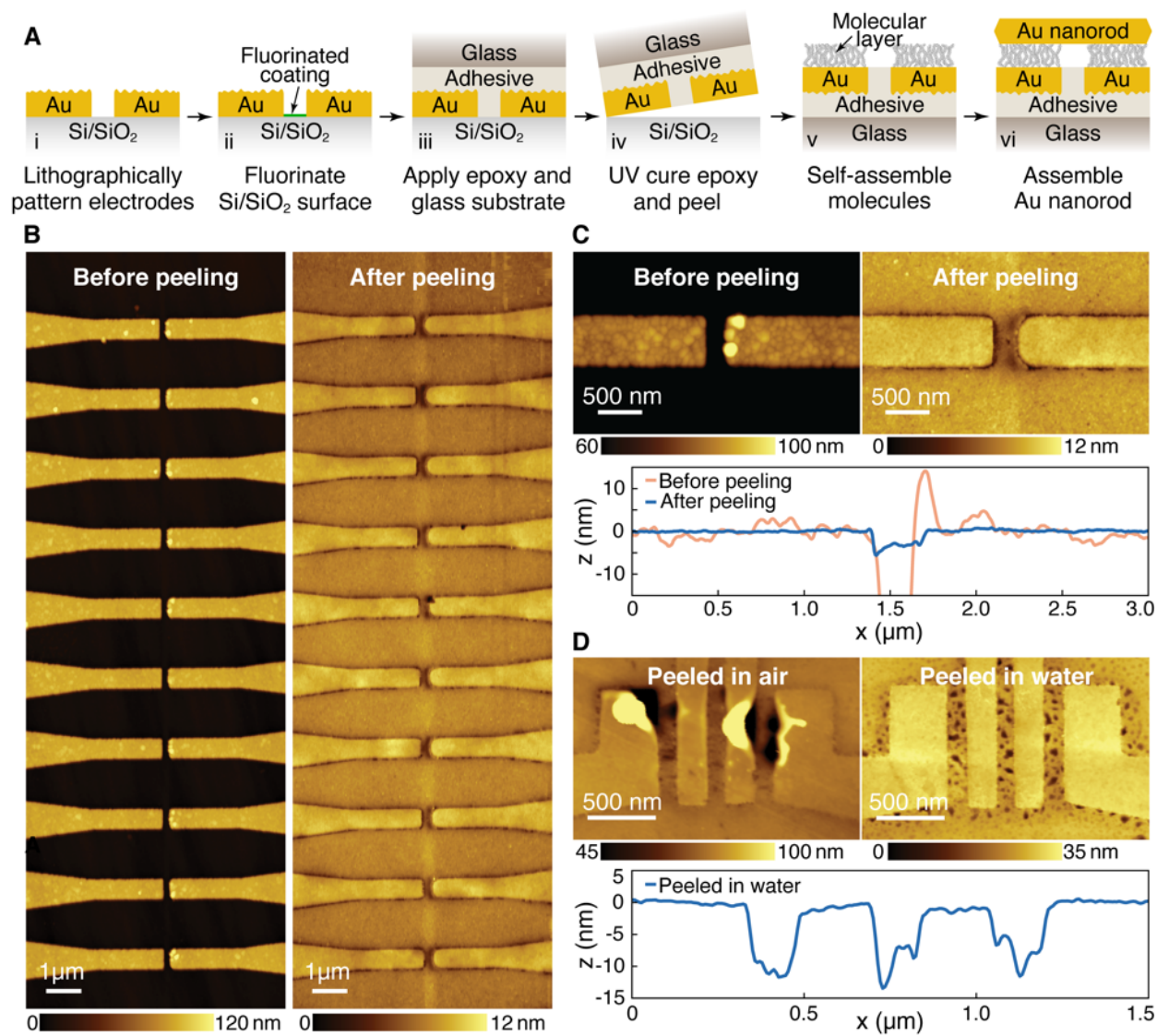

*Fig. 2. Ultrasmooth bottom electrodes. (A) Schematic of fabrication process – lithographically defined evaporated gold bottom electrodes are template-stripped to form ultrasmooth surfaces upon which a molecular layer is self-assembled. Subsequently, a colloidal gold nanorod is dielectrophoretically trapped to form a molecular junction. (B) AFM image of representative arrays of electrodes before and after planarization. (C) A magnified view of representative pairs of electrodes before and after planarization with the corresponding cross-section profiles. (D) Triboelectric damage is observed when bottom contacts are template-stripped in air and is rectified by submerging the sample in water during the process.*

In this approach, bottom contacts and the accompanying circuitry for electrical connections are first lithographically patterned on an atomically smooth silicon surface. Then, using an adhesive layer and a glass receiving substrate, the features are peeled from the silicon. The exposed surface of gold is much smoother with a roughness of ~ 0.6 nm rather than the ~ 2 nm roughness typical of as-evaporated gold. Atomic force microscope (AFM) images in Fig. S1 illustrate this improvement. The peeling process also planarizes the surface such that any edge imperfections resulting from the lift-off process used for their patterning are eliminated. Consequently, conventional methods such as lithography and thermal evaporation can be used in making the bottom contacts such that wafer scale processability is still accommodated while sub-nanometer surface uniformity is achieved. An example of arrays of two-terminal electrodes before and after peeling are shown in Fig. 2b. A magnified view in Fig. 2c captures the dramatically improved surface topography in more detail.

As device features are further miniaturized, we observe damage to the electrodes during the peeling process. We attribute this to a triboelectric effect causing accumulation of charges and damage as the two surfaces are separated from each other. This can be alleviated by performing the peeling while keeping the sample submerged in deionized (DI) water. Fig. 2d shows an example of a damaged structure and the outcome when alternatively peeled in water.

**Assembly of uniform molecular gaps**

Enabled by functional moieties, molecules can self-assemble on the bottom contacts with their structure defining the thickness and morphology of the resulting film. Amongst various assembly approaches, thiol-chemistry is widely used to self-assemble molecular monolayers on gold. Here, as an example, poly(ethylene glycol)thiol (PEG-thiol) is self-assembled on the bottom contacts in solution phase. Its uniform assembly can be characterized with high-resolution tip-enhanced near-field Fourier-transform infrared spectroscopy (nano-FTIR). An average nano-FTIR spectrum of PEG-thiol monolayer acquired on a peeled gold surface is shown in Fig. 3a. Molecules can alternatively or in combination be assembled on the top contact, providing more opportunities to tailor the gap functionality. As an example, a transmission electron microscope (TEM) image of a PEG-thiol functionalized gold nanorod which can be used as the top contact is shown in Fig. 3b.

The final step is the placement of the top contact without inducing damage to the underlying molecular layer. This, in particular, is challenging when the molecular layer has low packing density and low thermal stability such as the PEG-thiol used here. In such a scenario, a room temperature additive fabrication method such as dielectrophoretic assembly is suitable to create a uniform junction (15, 19, 20). The trapping process employs the laterally spaced bottom contacts to apply an AC voltage, inducing a trapping force which selectively attracts the rods dispersed in the surrounding solution to bridge the contacts. If conditions are optimized, the trapping can be limited to a single particle. However, it is likely to also trap multiple particles (Fig. S2). Parallel trapping can also be accommodated for scalable processing. Our design is shown in Fig. S3 and described in the Materials and Methods section. In our studies, we used nanorods of different aspect ratios, examples included in Fig. S4, but it should be noted that the process is applicable to other nanoparticles as well.

Fig. 3c shows an example of the final structure where a nanorod bridges the bottom contacts with self-assembled PEG-thiol on the surface. The resulting molecular gap formed between the ultrasmooth bottom gold surface and the flat facets of the nanorod can be uniform and well-defined as shown in Fig. 3d.

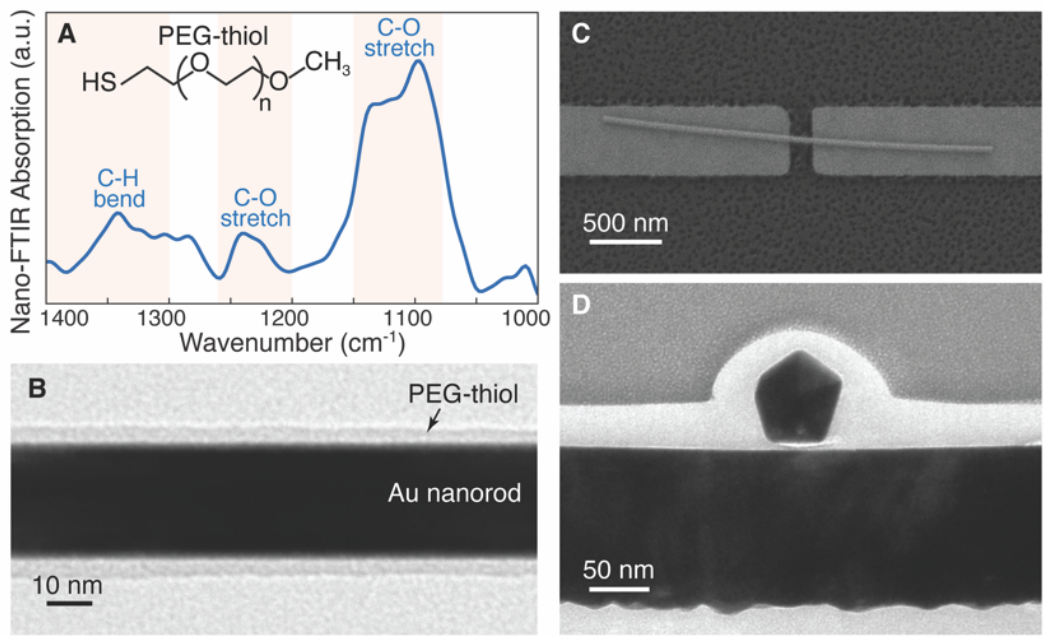

***Fig. 3. Molecular gap formation. (A)** nano-FTIR spectrum of PEG-thiol self-assembled on a peeled gold surface. **(B)** TEM image of a gold nanorod surface functionalized with PEG-thiol. **(C)** A top view scanning electron microscope (SEM) image of an assembled device. **(D)** TEM cross-section image of a uniform molecular gap.*

**Characterizing the active molecular gaps**

We fabricated devices with PEG-thiol and dodecanethiol gaps. The two molecules are selected since their corresponding self-assembled layers are expected to exhibit different thicknesses, morphologies and mechanical properties. Poly(ethylene glycol) monolayers are known to be compressible while dodecanethiol forms a packed film with a higher Young's modulus (21, 22). Representative current-voltage (I-V) characteristics of the devices are shown in Fig. 4. Evidently, the two devices perform differently. Within the applied voltage range, the PEG-thiol device undergoes a more abrupt change in the current and is accompanied by a larger hysteresis. Given that our design allows for the top contact to be mobile, we postulate that this accounts in part for the differences observed. We expect that the top contact can respond to an electrostatic force induced by the voltage applied to the bottom contacts such that the metal surfaces are attracted to each other while the molecular layer is compressed and the gap width reduced.

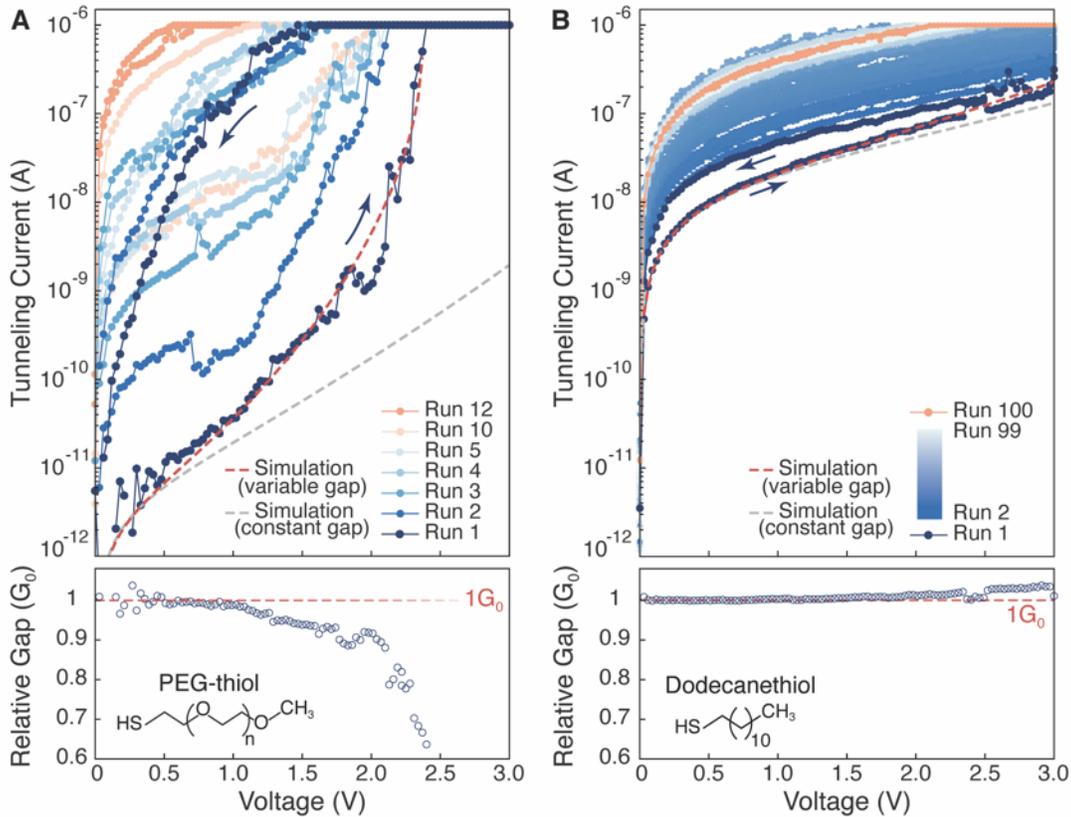

*Fig. 4. Current-voltage characteristics of molecular gaps. Electrical performance of a (A) PEG-thiol, and (B) dodecanethiol molecular junction over multiple runs with the arrows indicating sweep directions. For PEG-thiol, only select runs are plotted to allow for clear visualization of the performance progression over the tested runs. Experimental results are fitted employing the Simmons model considering a variable tunneling gap and is compared against the case when the gap remains fixed. The extracted relative change in the tunneling gap as a function of applied voltage is displayed where $G_0$ indicates the average initial gap thickness prior to actuation ($G_0$ = 2.5 nm and 1.5 nm for PEG-thiol and dodecanethiol, respectively). The negligible increase in the relative gap observed for the dodecanethiol example is within the bounds of the measurement error. The corresponding fitting parameters are: PEG-thiol device – $\Phi$ = 3.2 eV, $\alpha$ = 0.59, $\varepsilon_r$ = 2.6, Y = 81 MPa; Dodecanethiol device – $\Phi$ = 3.4 eV, $\alpha$ = 0.41, $\varepsilon_r$ = 2.5, Y = 0.97 GPa.*

To further investigate the characteristics, we fitted the experimental data against the simulated performance determined by the Simmons model while taking into account the possibility of mechanical deformation causing the tunneling gap to change during actuation. The Simmons model (eq. 1) has been widely used to describe the direct tunneling through a metal-molecule-metal structure (13, 23, 24):

$$I = \left(\frac{qA}{4\pi^2 \hbar G^2}\right)\left\{\left(\Phi - \frac{qV}{2}\right)\exp\left[-\frac{2(2m)^{1/2}}{\hbar}\alpha\left(\Phi - \frac{qV}{2}\right)^{1/2} G\right]\right.$$
$$\left.-\left(\Phi + \frac{qV}{2}\right)\exp\left[-\frac{2(2m)^{1/2}}{\hbar}\alpha\left(\Phi + \frac{qV}{2}\right)^{1/2} G\right]\right\} \quad (1)$$

where $m$ and $q$ are the electron mass and charge respectively, $A$ is the area of the tunneling junction, $\hbar$ is the reduced Plank constant, $V$ is the applied voltage, $G$ is the tunneling distance defined by the width of the nanogap, $\Phi$ is the tunneling barrier height, and $\alpha$ is an adjustable parameter that accounts for the electron effective mass and the barrier shape.

Based on the Simmons model, the tunneling current is exponentially dependent on the tunneling distance ($G$) which in our design is determined by the gap width. Thus, changes to the gap during device actuation can contribute to the extent of the current modulation. To take this into account in our simulation, we use the equilibrium equation of motion (eq. 2) to first evaluate the gap size at each applied voltage. We assume that the attractive electrostatic and van der Waals forces are balanced by the elastic restoring force provided by the molecular layer ($F_{eletrostatic} + F_{vdW} - F_{elastic} = 0$), expressed as:

$$\frac{\varepsilon_0 \varepsilon_r A V^2}{2G^2} + \frac{A_H A}{6\pi G^3} - k(L - G) = 0 \quad (2)$$

where $L$ is the thickness of the uncompressed molecular layer which will be larger than the initial gap thickness due to the presence of van de Waals forces, $\varepsilon_0$ is the permittivity of free space, $\varepsilon_r$ is the dielectric constant of the molecular layer, $A_H$ is the Hamaker constant taken to be $3\times10^{-19}$ J (25), and $k$ is the spring constant given by $k = YA/L$ where $Y$ is the Young's modulus of the molecular thin-film. This simplified model does not account for any nonlinearities in the mechanical behavior.

Once the tunneling gaps are evaluated, they are used with the Simmons model to simulate the expected current modulation. To converge on a suitable fit to the experimental results, the unknown parameters ($L$, $\Phi$, $\alpha$, $Y$, $\varepsilon_r$) are randomly selected from a pre-specified range and the simulation is iterated and parameters which minimize the fitting error are selected. The dotted red curves in Fig. 4 show the simulated behavior for the forward sweep of the first I-V run assuming a changing tunneling gap. Compared to the case where no compression of the molecular layer is considered, depicted by the dotted gray curves, a variable gap more closely captures the experimental I-V performance. Based on the best fits shown, Young's modulus of the PEG-thiol layer is inferred to be ~ 81 MPa while that of the dodecanethiol ~ 0.97 GPa, within the expected range of values experimentally measured and reported in the literature (21, 22). It should be noted though that given the large number of unknown parameters, it is not feasible to find a unique fit to the experimental results. Other combinations of the unknown parameters can also give a good fit, but the values are consistently within the physically expected bounds.

Using the $\Phi$ and $\alpha$ extracted from the fit, Simmons model is used to evaluate the change in the tunneling gap normalized to the average initial gap size (Fig. 4). Using the best fit parameters, the starting gap ($G_0$) for the PEG-thiol is estimated to be ~ 2.5 nm and that for the dodecanethiol ~ 1.5 nm. The PEG-thiol film shows ~ 37% compression within the 2 V applied voltage range, before reaching the set current compliance of the measurement tool, while a negligible compression is detected for the stiffer dodecanethiol gap. Overall, the experiment and its comparison to the simulated behavior suggest electromechanical reconfiguration of the tunneling gap with an extent dependent on the mechanical properties of the materials employed. It should also be noted, as shown in Fig. 4, that this behavior is accompanied by a gradual increase in the gap conductance throughout the subsequent measurement runs, and in some cases, leads to an eventual shorting of the device. Another example is shown in Fig. S5 where the positive and negative voltage sweeps are also included for reference. Such irreversible change may occur from damage to the molecular layer and/or the electrode surfaces, caused by current-induced local heating, metal from the electrodes electromigrating, or welding of the top and bottom electrodes at high voltages. The same degradation trend was not observed when the molecules were replaced with a structurally stable thin-film of hafnium oxide (Fig. S6).

In a design utilizing nanorods with a larger diameter (Fig. S4a), devices with hexanedithiol molecular gap were fabricated, an example of which is shown in Fig. 5a. The corresponding dark-field image of this device is included in Fig. S7. The representative I-V behavior of such a device (Fig. 5b) demonstrates stable performance with no hysteresis over consecutive runs and closely matches the simulated behavior based on the Simmons model. Once the applied voltage is increased beyond 2 V, the performance destabilizes (Fig. S8), and exhibits performance degradation similar to that shown in Fig. 4. As a result, it is evident that the testing/operating conditions for each device design plays a significant role in ensuring a stable performance.

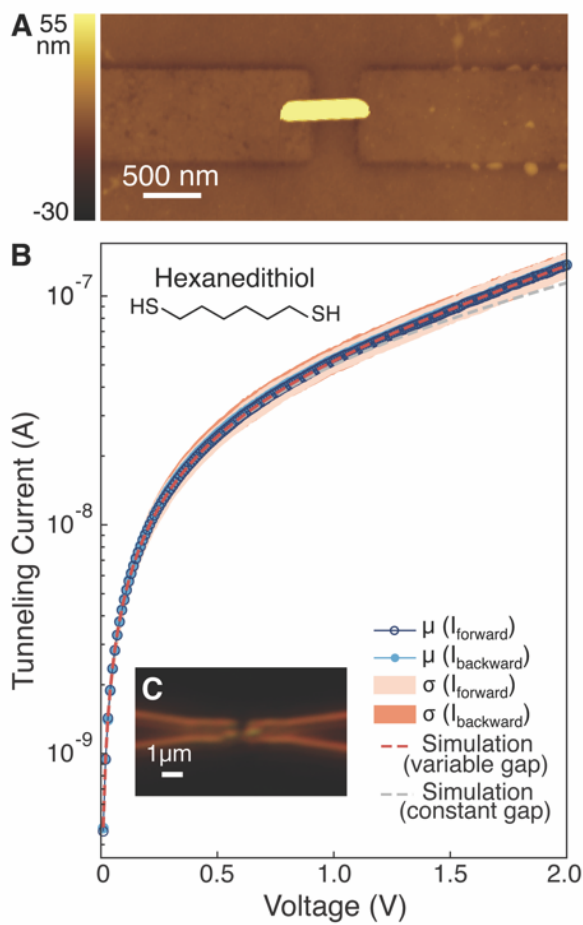

*Fig. 5. Optimization towards stable performance. (A) AFM image of a representative single rod device. (B) Current-voltage characteristic of a device with a hexanedithiol molecular layer averaged over 30 runs. The narrow standard deviation ($\sigma$) about the mean ($\mu$) and the lack of hysteresis suggest stable performance within the tested voltage range. Experimental results are fitted employing the Simmons model considering a variable tunneling gap and is compared against the case when the gap remains fixed ($G_0$ = 1.0 nm). The corresponding fitting parameters are: $\Phi$ = 3.7 eV, $\alpha$ = 0.42, $\varepsilon_r$ = 2.5, Y = 2.1 GPa. (C) Dark-field image of the device tested in (B).*

**Discussion**

The fabrication scheme introduced in this study provides a versatile platform to create < 5 nm uniform molecular gaps and can be readily applied to diverse material systems beyond those demonstrated here. The scalable planarization technique implemented for the bottom contacts can also be broadly employed for applications other than molecular devices in which surfaces with sub-nanometer uniformities are desired. The dielectrophoretic trapping of a colloidal nanorod allows room-temperature addition of the top contact with an atomically defined facet. However, even though parallel trapping for large area device processing is feasible, ensuring scalable single particle trapping remains a challenge and requires further optimization. For applications where this is critical, alternative methods such as aligned nanotransfer printing of the top contacts (12, 13) can also be used as it is compatible with the bottom electrode fabrication and does not damage the self-assembled molecules.

In our design, the thickness of the self-assembled molecular layer helps precisely define the width of the resulting nanogap and can lead to features with dimensions much smaller than lithographically feasible. These electrically accessible molecular gaps can be used to design and study active molecular junctions. Furthermore, these small nanogaps would also display functionalities beyond that introduced by the molecules themselves. For example, small dimensions can unlock unique interactions with light which can enable electrically-probed plasmonic structures.

We also show that the morphology and mechanical properties of the molecular layer may contribute to differences in the current modulation. The design allows the top contact to be mobile and mechanically respond to applied forces. Thus, when a voltage is applied to the bottom contacts, the induced electrostatic forces can attract the top contact towards the bottom. If the molecular layer is compressible, throughout this process the molecules may compress and reduce the nanogap width. As the gap decreases, the tunneling current can exponentially increase. Such electromechanical modulation of the nanogap can provide an additional control for engineering the current conduction. Fabrication and operation of such a reconfigurable nanogap, < 5 nm in size, is conventionally a challenge as large surface adhesive forces often cause structural instability and collapse. Presence of the molecules can help control these nanoscale forces, providing an opportunity for diverse applications in switches, sensors and actuators (26-30). It should however be noted that our current demonstration shows a degradation of device performance over multiple runs which may arise from degradation of the molecular/metal layers. Further studies are essential to investigate this observation and develop stable junctions by optimizing the materials and the operating conditions for the desired application.

**Materials and Methods**

**Device fabrication**

Ultrasmooth bottom electrodes: Bottom electrodes and the accompanying circuitry were first patterned on a silicon (or silicon oxide) substrate using electron beam lithography followed by gold thermal evaporation and lift-off. The sample was then surface functionalized with 1H,1H,2H,2H-perfluorooctyltriethoxysilane so that the silicon substrate assumes an anti-stiction property to allow for clean delamination of the receiving substrate upon cleavage. Vapor-phase growth of the fluorosilane was carried out for ~ 45 min in a vacuum desiccator. Then, ~ 30 μL of Norland Optical Adhesive (NOA 61) was deposited onto the substrate and a clean piece of glass was placed on top. The entire assembly was then placed in a vacuum desiccator for ~ 5 min to allow for the adhesive to spread evenly. Once evenly spread, the assembly was placed under UV irradiation to allow the adhesive to cure. Next, the substrate was annealed at 50 °C for 2 hours, and then the glass substrate along with the gold electrodes was peeled off the silicon. The peeling was initiated with the help of a blade and revealed the ultrasmooth underlying layer of the gold electrodes. For samples with smaller features, to avoid triboelectric damage, the peeling was performed with the sample submerged in DI water in a petri dish. The peeled sample was placed in a vial of ethanol and submerged in a sonicator for short bursts to clean the surface of any residual contaminants.

Molecular layer self-assembly: Poly(ethylene glycol) (PEG) with thiol termination was purchased from Creative PEGWorks. Alkanethiol molecules (dodecanethiol and hexanedithiol) were purchased from Sigma Aldrich. To self-assemble PEG-thiol, the peeled sample was placed in a solution of 5 mM PEG-thiol (5000 g/mol) in DI water. After 9-12 hours, the substrate was removed from the PEG-thiol solution, gently rinsed in DI water and dried with nitrogen. Dodecanethiol and hexanedithiol monolayers were formed by placing the peeled bottom electrodes in a 20 mM solution of the molecules in anhydrous ethanol in a nitrogen glovebox. After 2 hours, the samples were removed, rinsed in ethanol and dried with nitrogen. It should be noted that if the substrates are left in ethanol for too long, the surfaces become wrinkled.

Gold nanorod dielectrophoretic trapping: Gold nanorods were either purchased from Sigma Aldrich (716960-10ML) (Fig. S4a), or synthesized using a previously reported procedure (Fig. S4b) (31). First, the nanorod solutions were centrifuged twice to remove the excess surfactant. For large area trapping, upon centrifugation the rods were dispersed in mixed solution of DI water and ethanol. Then, ~ 10 µL of the rod solution was dispensed on top of the peeled bottom substrates. The ethanol/water mixture allows for the droplet to spread on the surface forming a thin layer. Then, using the trapping electrodes (Fig. S3), an AC voltage (typically 3-4 V at 1 MHz) was applied for ~ 2 min. The voltage was then reduced to ~ 0.5 V and the sample was left to dry. To further promote single particle trapping such as the device in Fig. 5a, a variation of the above technique was employed. Upon centrifugation, the nanorods (Fig. S4a) were diluted in DI water. Then, the nanorod solution was dispensed with the aid of a pre-pulled microcapillary (purchased from World Precision Instruments Inc.) and a syringe pump such that the dispensed volume is small and controlled. The trapping was then performed with an AC voltage (typically 3 V at 3 MHz) applied for ~ 5 min or until sample was dried. Successful trapping was confirmed by dark-field optical microscopy and atomic force microscopy.

**Device characterization**

Once successful dielectrophoretic assembly of the top electrode was confirmed, the device electrical characterization was performed using a B1500A Agilent Semiconductor Parameter Analyzer in air. The cross-sectional TEM sample preparation was performed on an FEI Helios 660 Dual-beam SEM-FIB. First the sample was sputter-coated with 10-50 nm of carbon. The lamella then was prepared by coating the device region locally with electron-beam deposited platinum, followed by ion-beam deposited platinum. The sample was ion-beam milled and extracted with an Omniprobe attachment and attached to a copper TEM half-grid. The lamella was center-mounted on a wide post, and then thinned to a final thickness of ~ 40 nm. TEM imaging was performed with a JEOL 2100 TEM at 200 kV with a double-tilt sample holder. Nano-FTIR spectroscopy was performed with a neaSNOM scattering type near-field optical microscope (neaspec GmbH) (32, 33). The nano-FTIR absorption spectra were recorded with a spectral resolution of ~ 10 cm$^{-1}$ and co-averaging of 25 individual spectra. Each sample spectrum is normalized to a reference spectrum measured on a clean gold sample surface.

**Device modelling and fitting of the experimental results**

The experimental results were fitted using the Simmons tunneling model (eq. 1) taking into account the potential change in the molecular gap due to the electrostatic force induced by the applied voltage. In this analysis, the equilibrium equation of motion (eq. 2) was first solved to determine the tunneling gap at each applied voltage. Then, the extracted values were used in the Simmons model to simulate the expected current modulation. The tunneling barrier height ($\Phi$), tunneling barrier shape ($\alpha$), Young's modulus ($Y$), dielectric constant ($\varepsilon_r$), and the length of the uncompressed molecular layer ($L$) are unknown fitting parameters. To gain an understanding of the experimental data with respect to the simulated device behavior, the model was executed over a large number of runs using randomly selected combinations of parameters within a predefined range and the case leading to the closest fit to the experiment was determined. Given the large number of unknown parameters, acquiring a unique fit to the experimental data is not feasible and various sets of parameters can lead to a good fit. However, these values are within the physically expected range, consistent with those experimentally measured and reported in the literature. A more comprehensive analysis would require additional metrology methodologies to independently measure the unknown parameters or the motion.

**References**

1. N. Xin, J. Guan, C. Zhou, X. Chen, C. Gu, Y. Li, M. A. Ratner, A. Nitzan, J. F. Stoddart, X. Guo, Concepts in the design and engineering of single-molecule electronic devices. *Nat. Rev. Phys.* **1**, 211-230 (2019).
2. S. Aradhya, L. Venkataraman, Single-molecule junctions beyond electronic transport. *Nat. Nanotech.* **8**, 339-410 (2013).
3. A. Vilan, D. Aswal, D. Cahen, Large-area, ensemble molecular electronics: motivation and challenges. *Chem. Rev.* **117**, 4248-4286 (2017).
4. H. Jeong, D. Kim, D. Xiang, T. Lee, High-yield functional molecular electronic devices. *ACS Nano* **11**, 6511-6548 (2017).
5. R. T. Hill, J. J. Mock, A. Hucknall, S. D. Wolter, N. M. Jokerst, D. R. Smith, A. Chilkoti, Plasmonic ruler with angstrom length resolution. *ACS Nano* **2012**, 9237-9246 (2012).
6. G. Li, Q. Zhang, S. A. Maier, D. Lei, Plasmonic particle-on-film nanocavities: a versatile platform for plasmon-enhanced spectroscopy and photochemistry. *Nanophotonics* **7**, 1865-1889 (2018).
7. H. Qian, S. Hsu, K. Gurunatha, C. T. Riley, J. Zhao, D. Lu, A. R. Tao, Z. Liu, Efficient light generation from enhanced inelastic electron tunnelling. *Nat. Photon.* **12**, 485-488 (2018).

8. C. Zhou, M. R. Deshpande, M. A. Reed, L. Jones, J. M. Tour, Nanoscale metal/self-assembled monolayer/metal heterostructures. *Appl. Phys. Lett.* **71**, 611-613 (1997).
9. H. B. Akkerman, P. W. M. Blom, D. M. De Leeuw, B. De Boer, Towards molecular electronics with large-area molecular junctions, *Nature* **441**, 69-72 (2006).
10. G. Puebla-Hellmann, K. Venkatesan, M. Mayor, E. Lörtscher, Metallic nanoparticle contacts for high-yield, ambient-stable molecular-monolayer devices. *Nature* **559**, 232-235 (2018).
11. Y. Yang, C. Gu, J. Li, Sub-5 nm metal nanogaps: physical properties, fabrication methods, and device applications. *Small* **15**, 1804177 (2019).
12. Y. L. Loo, D. V. Lang, J. A. Rogers, J. W. P. Hsu, Electrical contacts to molecular layers by nanotransfer printing. *Nano Lett.* **3**, 913-917 (2003).
13. J. R. Niskala, W. C. Rice, R. C. Bruce, T. J. Merkel, F. Tsui, W. You, Tunneling characteristics of Au-alkanedithiol-Au junctions formed via nanotransfer printing (nTP). *J. Am. Chem. Soc.* **134**, 12072-12082 (2012).
14. V. Dubois, S. N. Raja, P. Gehring, S. Caneva, H. S. J. van der Zant, F. Niklaus, G. Stemme, Massively parallel fabrication of crack-defined gold break junctions featuring sub-3 nm gaps for molecular devices. *Nat. Commun.* **9**, 3433 (2018).
15. J. A. Fereiro, X. Yu, I. Pecht, M. Sheves, J. C. Cuevas, D. Cahen, Tunneling explains efficient electron transport via protein junctions. *PNAS* **115**, E4577-E4583 (2018).
16. E. A. Weiss, R. C. Chiechi, G. K. Kaufman, J. K. Kriebel, Z. Li, M. Duati, M. A. Rampi, G. M. Whitesides, Influence of defects on the electrical characteristics of mercury-drop junctions: self-assembled monolayers of n-alkanethiolates on rough and smooth silver. *JACS* **129**, 4336-4349 (2007).
17. L. Yuan, L. Jiang, D. Thompson, C. A. Nijhuis, On the remarkable role of surface topography of the bottom electrodes in blocking leakage current in molecular diodes. *JACS* **136**, 6554-6557 (2014).
18. E. A. Weiss, G. K. Kaufman, J. K. Kriebel, Z. Li, R. Schalek, G. M. Whitesides, Si/SiO$_2$-templated formation of ultraflat metal surfaces on glass, polymer, and solder supports: their use as substrates for self-assembled monolayers. *Langmuir* **23**, 9686-9694 (2007).
19. A. Barik, X. Chen, S. H. Oh, Ultralow-power electronic trapping of nanoparticles with sub-10 nm gold nanogap electrodes. *Nano Lett.* **16**, 6317-6324 (2016).
20. E. M. Freer, O. Grachev, X. Duan, S. Martin, D. P. Stumbo, High-yield self-limiting single-nanowire assembly with dielectrophoresis. *Nat. Nanotech.* **5**, 525-530 (2010).
21. Q. Huang, I. Yoon, J. Villanueva, K. Kim, D. J. Sirbuly, Quantitative mechanical analysis of thin compressible polymer monolayers on oxide surfaces. *Soft Matter* **10**, 8001-8010 (2014).
22. F. W. DelRio, C. Jaye, D. A. Fischer, R. F. Cook, Elastic and adhesive properties of alkanethiol self-assembled monolayers on gold. *Appl. Phys. Lett.* **94**, 131909 (2009).
23. J. G. Simmons, Generalized formula for the electric tunnel effect between similar electrodes separated by a thin insulating film. *J. Appl. Phys.* **34**, 1793-1803 (1963).
24. W. Wenyong, T. Lee, M. A. Reed. Mechanism of electron conduction in self-assembled alkanethiol monolayer devices. *Phys. Rev. B* **68**, 035416 (2003).
25. G. Palasantzas, P. J. van Zwol, J. Th. M. De Hosson, Transition from Casimir to van der Waals force between macroscopic bodies. *Appl. Phys. Lett.* **93**, 121912 (2008).

26. F. Niroui, A. I. Wang, E. M. Sletten, Y. Song, J. Kong, E. Yablonovitch, T. M. Swager, J. H. Lang, V. Bulović, Tunneling nanoelectromechanical switches based on compressible molecular thin films. *ACS Nano* **9**, 7886-7894 (2015).
27. J. O. Lee, Y. H. Song, M. W. Kim, M. H. Kang, J. S. Oh, H. H. Yang, J. B. Yoon, A sub-1-volt nanoelectromechanical switching device. *Nat. Nanotech.* **8**, 36-40 (2013).
28. P. Jialong, H. Jeong, Q. Lin, S. Cormier, H. Liang, M. F. L. De Volder, S. Vignolini, J. J. Baumberg, Scalable electrochromic nanopixels using plasmonics. *Sci. Adv.* **5**, eaaw2205 (2019).
29. S. Caneva, P. Gehring, V. M. Carcía-Suárez, A. García-Fuente, D. Stefani, I. J. Olavarria-Contreras, J. Ferrer, C. Dekker, H. S. J. van der Zant, Mechanically controlled quantum interference in graphene break junctions. *Nat. Nanotech.* **13**, 1126-1131 (2018).
30. Q. Huang, J. Lee, F. T. Arce, I. Yoon, P. Angsantikul, J. Liu, Y. Shi, J. Villaneuva, S. Thamphiwatana, X. Ma, L. Zhang, S. Chen, R. Lal, D. J. Sirbuly, Nanofibre optic force transducers with sub-piconewton resolution via near-field plasmon-dielectric interactions. *Nat. Photonics* **11**, 352-355 (2017).
31. Y. N. Wang, W. T. Wei, C. W. Yang, M. H. Huang, Seed-mediated growth of ultralong gold nanorods and nanowires with a wide range of length tunability. *Langmuir* **29**, 10491-10497 (2013).
32. Amenabar, S. Poly, M. Goikoetxea, W. Nuansing, P. Lasch, R. Hillenbrand, Hyperspectral infrared nanoimaging of organic samples based on Fourier transform infrared nanospectroscopy. *Nat. Commun.* **8**, 14402 (2017).
33. F. Huth, A. Govyadinov, S. Amarie, W. Nuansing, F. Keilmann, R. Hillenbrand, Nano-FTIR absorption spectroscopy of molecular fingerprints at 20 nm spatial resolution. *Nano Lett.* **12**, 3973–3978 (2012).

**Acknowledgments**

We thank Harvard University Center for Nanoscale Systems research staff A. Akey and S. Kramer for FIB sample preparation assistance, and J. Gardener for TEM imaging assistance. We also thank T. Gokus of neaspec GmbH for assistance in carrying out the nano-FTIR measurements. **Funding:** This work was supported by the National Science Foundation (NSF) Center for Energy Efficient Electronics Science ($E^3S$) Award ECCS-0939514. MS and JP acknowledge support from the Natural Sciences and Engineering Research Council of Canada Postgraduate Scholarship. MS also acknowledges support from the NSF Graduate Research Fellowship Program under Grant No. 174530. This work was performed in part at the Harvard University Center for Nanoscale Systems (CNS), a member of the National Nanotechnology Coordinated Infrastructure Network (NNCI), which is supported by the National Science Foundation under NSF ECCS award no. 1541959. **Author Contributions:** FN developed the fabrication technique and device design. FN, MS and JH carried out the experiments including device fabrication and characterization. JP performed the FIB cross-sectional imaging. FN processed the experimental results and modeling. All authors discussed the results and contributed to editing the manuscript. **Competing interests:** A patent related to the template-stripping process presented here has been filed. **Data and materials availability:** All data needed to evaluate the conclusions in the paper are present in the paper and/or the Supplementary Materials. Additional data related to this paper may be requested from the authors.

# Supplementary Materials

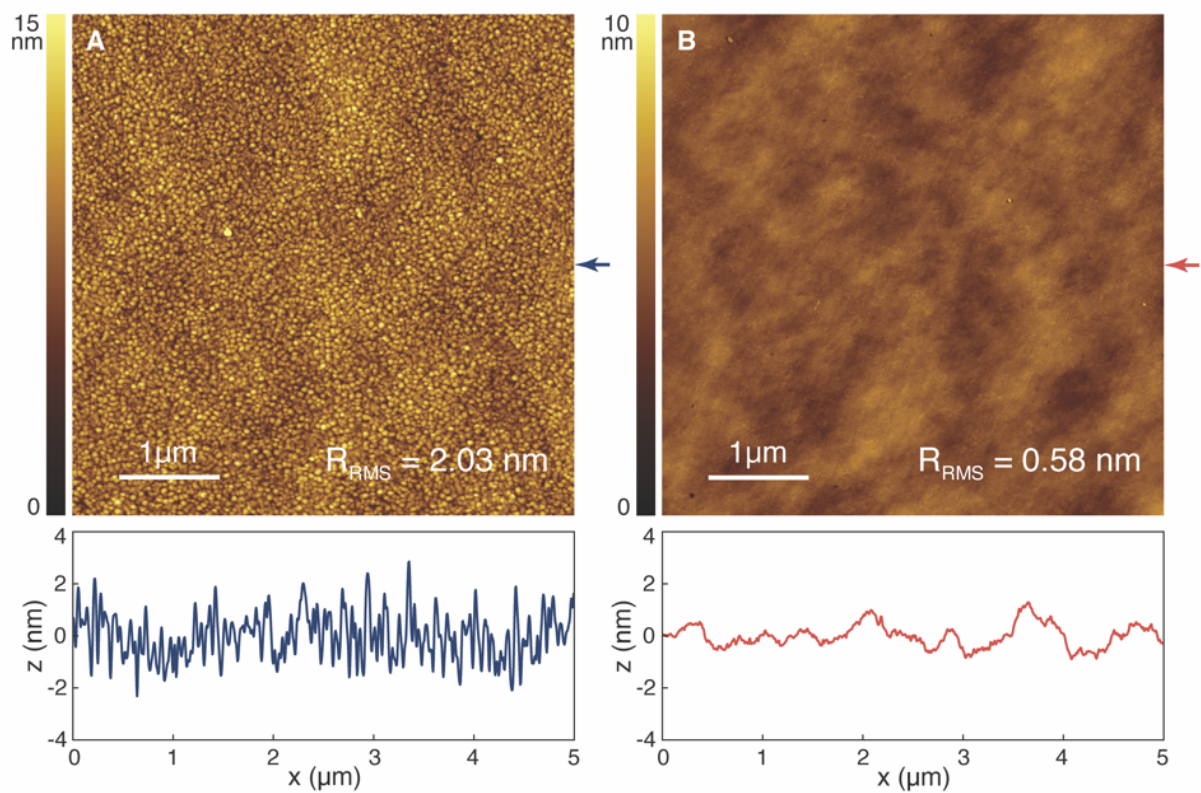

**Fig. S1. Surface roughness comparison.** AFM images of **(A)** as-evaporated and **(B)** templated-stripped gold and the corresponding surface profiles showing the reduction of surface roughness to < 1 nm.

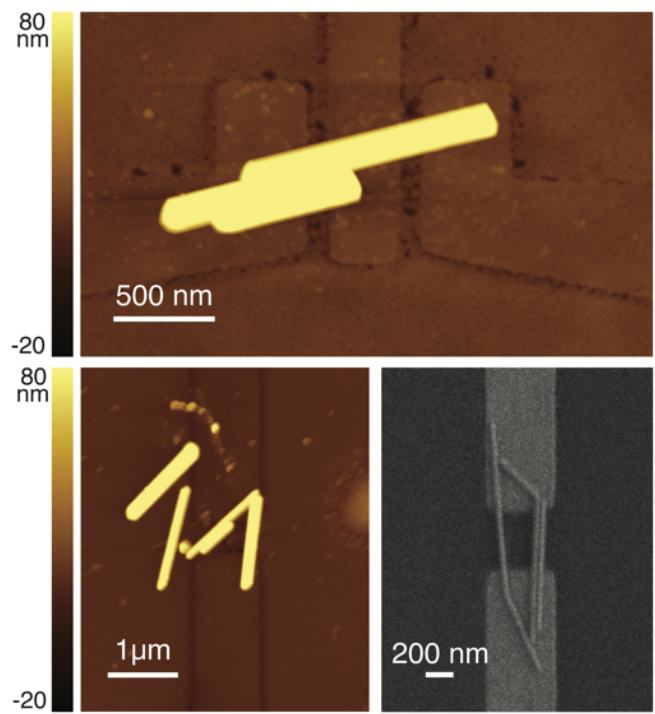

**Fig. S2. Trapping of multiple rods.** It is common to observe multiple rods trap during the dielectrophoretic assembly of the top contact as shown in the AFM and SEM images.

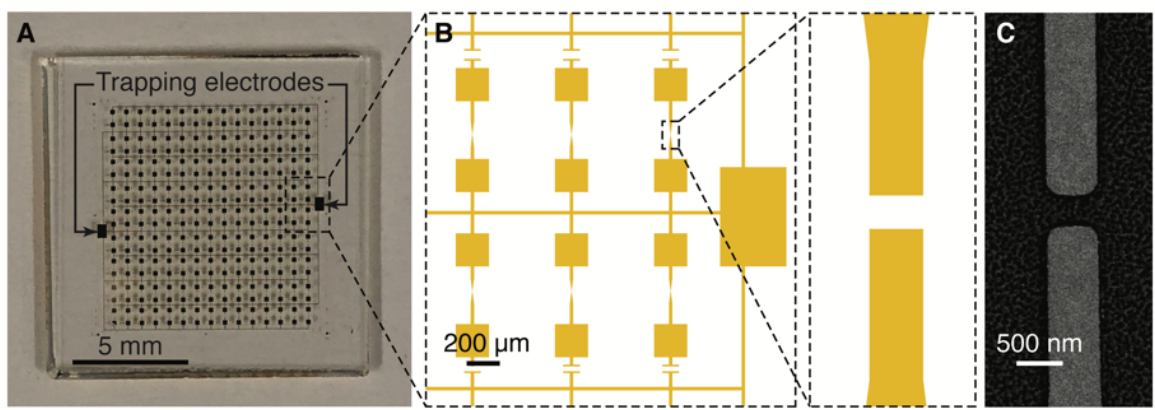

**Fig. S3. Bottom substrate layout.** **(A)** Photograph of a substrate with completed bottom electrodes showing the large area compatibility of the peeling process. **(B)** Schematic view of the layout showing the bottom contacts and the accompanying circuitry used for trapping and testing. The capacitors are incorporated in the design to allow for large area trapping of the nanorods using the trapping electrodes when an AC voltage is applied, while keeping the devices isolated from each other when individually characterized using their local electrodes. **(C)** SEM image of an example pair of bottom electrodes.

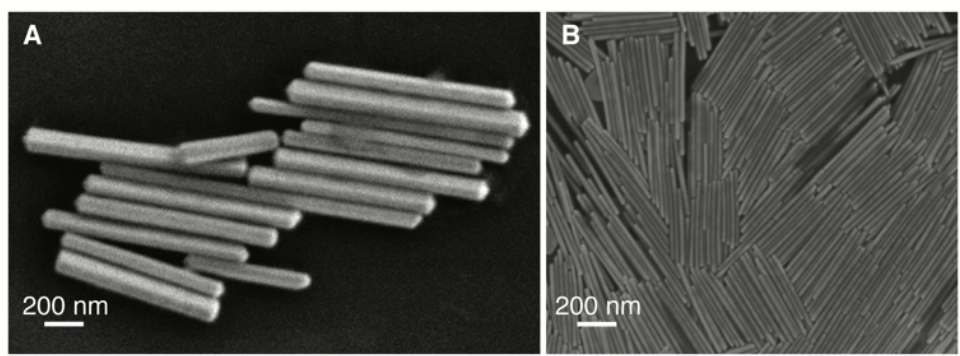

**Fig. S4. Gold nanorods.** SEM images of gold nanorods with different aspect ratios used for device fabrication.

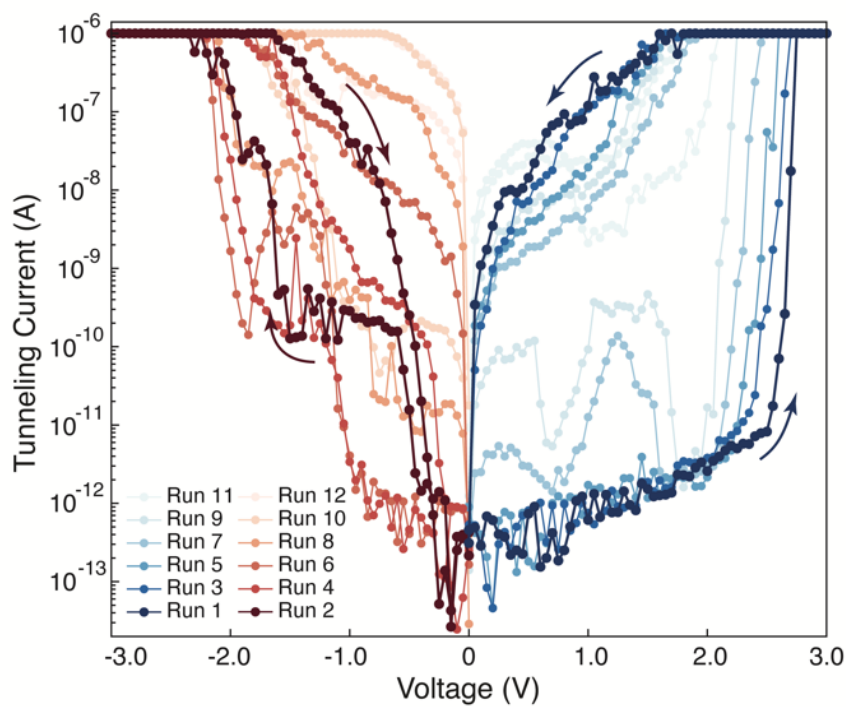

**Fig. S5. Positive and negative voltage sweeps.** Current-voltage characteristics of a PEG-thiol molecular junction tested over multiple runs alternating between positive and negative operating voltages. The arrows show the sweep directions and absolute values of the measured currents are displayed.

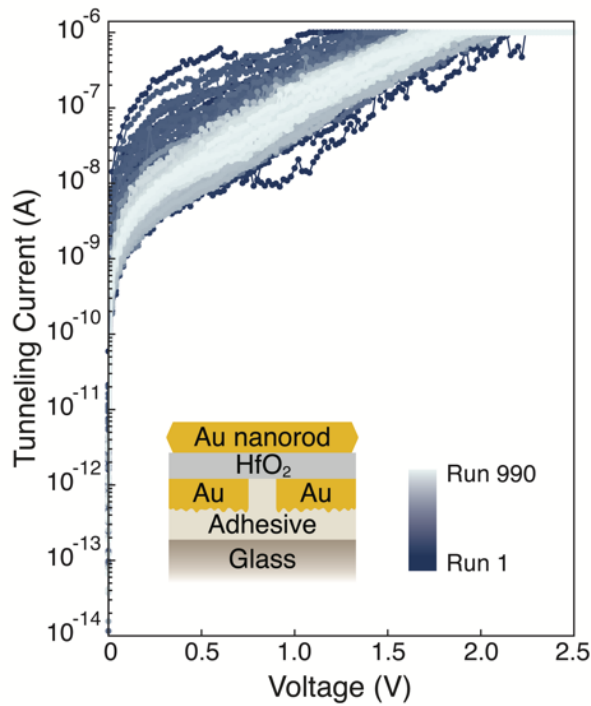

**Fig. S6. A hafnium oxide nanogap.** Current-voltage characteristics of a device with a hafnium oxide spacer layer deposited using atomic layer deposition, shown over 990 runs with every 10th run displayed. Initial runs show a larger variability and hysteresis but the device settles into a stable operation over the course of the measurement. The initial instability might be due to the degradation of the residual surfactant surrounding the nanorod.

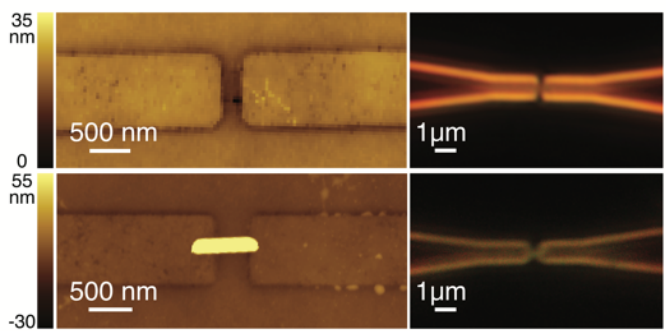

**Fig. S7. Examples of devices with and without a top contact.** Representative AFM and darkfield images of bottom contacts before and after trapping. Darkfield imaging allows high throughput confirmation of single rod placement for device characterization.

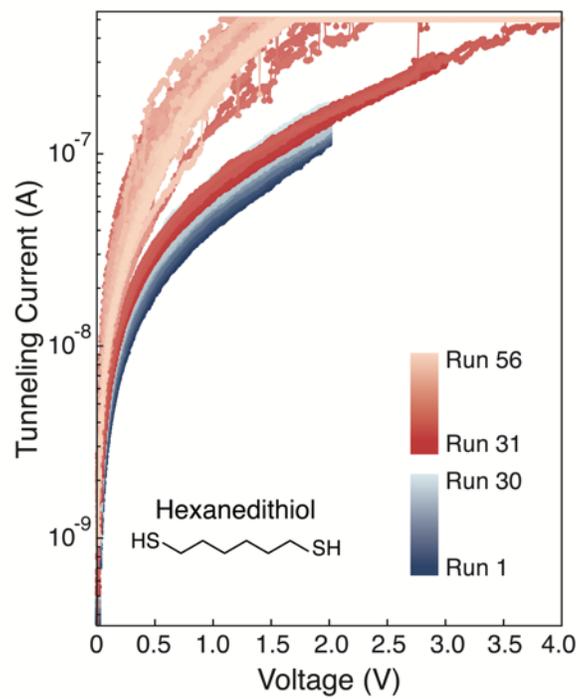

**Fig. S8. Performance degradation at high voltages.** Current-voltage characteristics of the device shown in Fig. 5 displaying performance degradation once optimal operating voltages are exceeded.